\documentclass[showpacs, twocolumn,preprintnumbers,amsmath,amssymb,amsfonts,superscriptaddress]{revtex4-1}
\usepackage{amsmath}
\usepackage{graphicx}
\usepackage{dcolumn}
\usepackage{bbm}
\usepackage{subfigure}
\usepackage{amssymb}
\usepackage{gensymb}
\usepackage{latexsym}
\usepackage{makecell}
\usepackage{footnote}
\usepackage{soul}
\usepackage{bm}
\usepackage{xcolor}
\usepackage{comment}
\usepackage[colorlinks=true,
            citecolor=blue,
            linkcolor=blue,
            urlcolor=blue]{hyperref}

\usepackage{tabularx}
\newcolumntype{P}[1]{>{\centering\arraybackslash}p{#1}}

\begin{document}

\title{Intrinsic Nernst Effect from Berry Curvature in Superconductors}

\author{Tzu-Chi~Hsieh}\affiliation{Department of Physics, University of Notre Dame, Notre Dame, Indiana 46556, USA}
\author{Cong Xiao}\affiliation{Interdisciplinary Center for Theoretical Physics and Information Sciences (ICTPIS), Fudan University, Shanghai 200433, China}
\author{Yi-Ting~Hsu}\affiliation{Department of Physics, University of Notre Dame, Notre Dame, Indiana 46556, USA}

\date{\today}

\begin{abstract}
The Nernst effect in superconductors is typically linked to fluctuating Cooper pairs above $T_c$ or vortex motion below $T_c$. We show instead that Berry curvature of Bogoliubov quasiparticles can generate an intrinsic Nernst response in a clean, vortex-free superconducting state. Focusing on two-dimensional systems with Ising spin–orbit coupling, relevant to transition-metal dichalcogenides, we identify two regimes: an intervalley 
$s$-wave paired state where a weak magnetic field activates the effect, and an intravalley chiral $p$-wave paired state that exhibits a \textit{spontaneous} charge or spin Nernst response without a field. We propose an experimental setup that circumvents screening and provide estimates of the signal magnitude. Our results establish the Nernst effect as a direct probe of Berry curvature and pairing symmetry in two-dimensional spin-orbit-coupled superconductors.
\end{abstract}

\maketitle

\twocolumngrid
 
\textit{Introduction---}
Unambiguous identification of topological superconductors (TSCs) remains a major challenge in the study of topological materials. 
In contrast to topological insulators and semimetals, where boundary modes provide clear experimental signatures \cite{Xia2009,Roushan2009,Xu2015TaP}, 
the detection of Majorana boundary modes in TSCs is often obscured by similar signals from topologically trivial mechanisms \cite{Yu2021NonMajorana,Jack2021Detecting,Mandal2023Topological}. 
To move beyond this limitation, it is crucial to broaden the evidence: 
signatures that appear \textit{away from the boundary} should complement boundary probes to conclusively identify TSCs.  
In two-dimensional (2D) chiral TSCs, quantized thermal Hall conductance is known as a direct probe of non-zero Chern numbers \cite{read_2000_paired}. In particular, 
the effective $p+ip$ TSC phase and its trivial counterpart in a proximitized Rashba 2D electron gas were predicted to exhibit distinct spectral and tunneling signatures \cite{liao_2025_detecting}.  

Here, we propose that quasiparticle thermoelectric effects originating from \textit{quasiparticle Berry curvature} provide an effective yet largely unexplored means to discriminate between topological and conventional superconductivity. 
Thermoelectric effects in superconductors have long been an interesting yet subtle topic \cite{ginzburg_1944_thermoelectric,harlingen_1982_thermoelectric,ginzburg_2004_nobel,Matsushita2022SpinNernstTSC,Matsushita2025SpinCaloritronicsNUSC,Matsushita2025IntrinsicSNE}: 
while quasiparticles at finite temperatures can in principle generate a thermoelectric signal, 
any intrinsic quasiparticle thermoelectric charge response is generally believed to be unobservable in standard bar geometries due to the screening effect. Experimentally, the transverse thermoelectric response, or Nernst effect, observed in superconductors has therefore been largely attributed to vortex motion below $T_c$ and to short-lived Cooper pairs above $T_c$ \cite{behnia_2016_nernst,song_2024_unconventional}. 
The intrinsic quasiparticle charge Nernst effect arising from Berry curvature has received little experimental and theoretical attention.

In the following, we will show that an intrinsic Nernst effect in superconductors can be observed even without magnetic vortices, and can distinguish topological from trivial superconductors. We propose a ring geometry with a radial temperature gradient as the ideal setup for detecting this effect [Fig.~\ref{fig1}(a)], where it manifests as a non-quantized magnetic flux at the ring’s center—a direct and measurable signature of quasiparticle-origin thermoelectricity. By performing semiclassical wave-packet analysis on a two-valley system with Ising and Rashba spin-orbit couplings, we demonstrate finite Nernst signals for (1) $s$-wave intervalley pairing at $B\neq0$, and (2) chiral $p$-wave intravalley pairing at $B=0$ [Fig.~\ref{fig1}(b,c)].   
Our work uncovers an overlooked mechanism for the transverse thermoelectric effect in superconductors, establishes a realistic ring-geometry setup and material design principles for its detection, and identifies a distinctive signature of topological superconductivity that does not rely on Majorana modes. The proposal applies broadly to quasi-2D superconductors with Rashba and Ising spin-orbit couplings, including monolayer and few-layer transition-metal dichalcogenides \cite{hsu_2017_topological,Hu2016Stacking,Zhang2021SpinTextured,Devakul2021Magic,Tian2024TowardDirect}.

\begin{figure}[h]
\includegraphics[width=.4\textwidth]{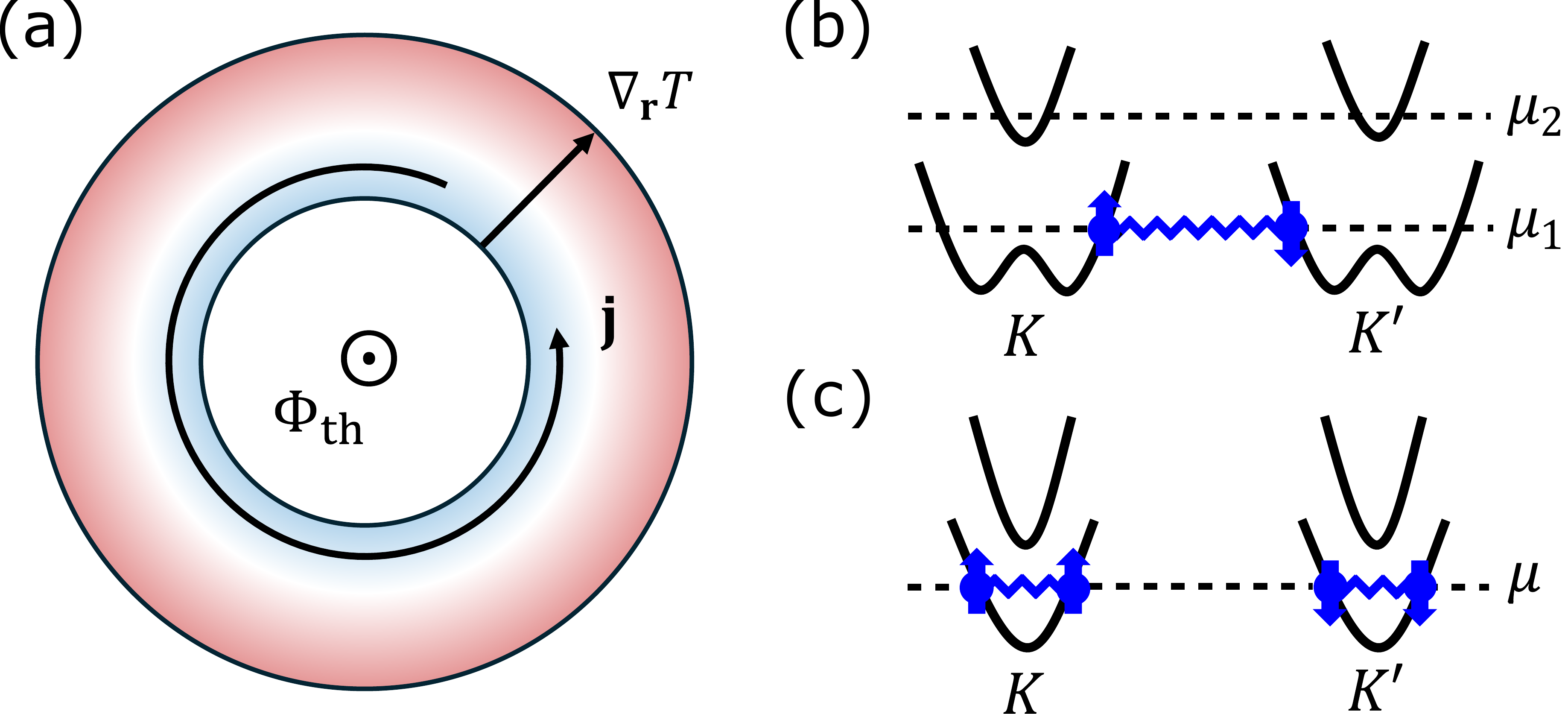}
\caption{(a) Schematic of the proposed setup 
for measuring Nernst effect in superconductors. A radial thermal gradient $\nabla_{\bf r} T\neq 0$ induces a transverse loop current ${\bf j}$, which generates a detectable out-of-plane magnetic flux $\Phi_{\rm th}$. The two types of paired states we consider in Ising superconductors:  the (b) intervalley $s$-wave and (c) intravalley $p$-wave pairings. $K$ and $K'$ label the valleys, and $\mu$ is the chemical potential with subscripts labeling different doping regimes. The blue arrows and wiggly lines denote the electron spins and pairing interactions, respectively.  \label{fig1}}
\end{figure}

\textit{Semiclassical wavepacket approach---\label{sec:WP approach}}
We employ a semiclassical wavepacket framework to capture the thermoelectric response in superconductors. 
Because both quasiparticles and Cooper pairs contribute to charge transport, a two-fluid description is needed to account for temperature-gradient-induced magnetic flux. 
A recent formulation of the semiclassical approach properly incorporates intrinsic transport effects in systems with non-conserved quasiparticle currents and is applicable to superconductors \cite{XiaoPRB2021}. 
Here, we use this framework to examine the Berry-curvature-induced Nernst conductance of Bogoliubov quasiparticles, with the counterflow of Cooper pairs self-consistently included \cite{DongPRL2020,XiaoPRB2021}.  

In superconductors, the condensate acts as a charge reservoir, so the quasiparticle charge current \cite{Nambu1960Superconductivity,Blonder1982,Nambu2009,Liang2017} 
${\bf j}_{qp}=\tfrac{1}{2}\{{\bf v},\rho\}$ is not conserved, where ${\bf v}$ and $\rho$ are the velocity and charge operators \cite{Ahn2021,velocity_operator}. 
In the Nambu basis, $\rho=\sigma_z$, with $\sigma_i$ ($i=x,y,z$) the particle–hole Pauli matrices. 
In the steady state, $\partial_t\sigma_z=0$, and the continuity equation becomes \cite{Shi2006ProperDefinitionSpinCurrent,Murakami2006QuantumSpinHall,ParameswaranPRL2012}
\begin{align}\label{eq:qp_continuity}
    \nabla_{\bf r}\!\cdot{\bf j}_{qp} = \mathcal{B}_p,
\end{align}
where the source term $\mathcal{B}_p = \tfrac{i}{\hbar}[H_p,\sigma_z]$ originates from the pairing Hamiltonian $H_p$ and satisfies $\mathcal{B}_p=-\nabla_{\bf r}\!\cdot{\bf j}_s$, representing the Cooper-pair counterflow \cite{ParameswaranPRL2012}. 
Thus, the total current ${\bf j}={\bf j}_s+{\bf j}_{qp}$ is conserved. 
Averaging Eq.~(\ref{eq:qp_continuity}) over the wavepacket and distribution function yields the semiclassical transport equation. 
To second order in gradient expansion, the linear response to temperature gradient was found to be ${\bf j}=\boldsymbol{\alpha}\cdot\nabla_{\bf r}T$ \cite{XiaoPRB2021}, where the Hall component of $\boldsymbol{\alpha}$ reads \cite{SM}
\begin{align}\label{eq:alpha_H}
    \alpha_{H} = \frac{e}{\hbar}\sum_n\int\frac{d^2k}{(2\pi)^2}\frac{dg_{n{\bf k}}}{dT}\Omega_{n{\bf k}}\rho_{n{\bf k}}. 
\end{align} 
Here, $e>0$ is the elementary charge, $\hbar$ the reduced Planck constant, $n$ the band index, ${\bf k}$ the momentum relative to the $K$ or $K'$ valley. $\Omega_{n{\bf k}}$ denotes the Berry curvature, $\rho_{n{\bf k}}=\langle\psi_{n{\bf k}}|\sigma_z|\psi_{n{\bf k}}\rangle$ and 
$g_{n{\bf k}}=-k_BT\ln(1+e^{-\beta E_{n{\bf k}}})$ with $\beta=1/k_BT$ are the state-resolved charge and grand potential, respectively. $E_{n{\bf k}}$ and $|\psi_{n{\bf k}}\rangle$ are the eigenenergy and eigenstate of the Bogoliubov–de Gennes (BdG) Hamiltonian. Equation~(\ref{eq:alpha_H}) applies to strong type-II superconductors such as 2D materials, where the penetration depth exceeds the system size.

\textit{Model---\label{sec:model}} We consider a mean-field model generic for 2D hexagonal superconductors with both Rashba and Ising spin-orbit couplings (SOC), representative of few-layer Van der Waals systems such as 2H-type TMDs. 
The normal-state Hamiltonian is \cite{Yuan2014,Zhou2019SVHE}
\begin{align}
    h_{0,\tau} = \xi_k + \lambda_{so}\hat{z}\cdot{\bf s}\times{\bf k} + \tau\beta_{so}s_z,
\end{align}
where momentum $k$ is measured from valley centers $K$ and $K'$ ($\tau=\pm1$). 
Here $s_{i=x,y,z}$ are spin Pauli matrices, and $\xi_k = tk^2 - \mu$ with hopping $t$ and chemical potential $\mu$. 
The Rashba SOC $\lambda_{so}$ and Ising SOC $\beta_{so}$ arise from broken out-of-plane mirror and in-plane inversion symmetries, respectively.

We study two distinct superconducting states described by the mean-field Hamiltonian 
$H=\tfrac{1}{2}\sum_{\tau=K,K'}\int_{\bf k}\Psi_{\tau}^\dagger ({\bf k})h_{\rm BdG,\tau}\Psi_{\tau}({\bf k})$,  
with Nambu basis $\Psi_{\tau}({\bf k})=[c_{\tau,\textbf{k},\uparrow},c_{\tau,\textbf{k},\downarrow},c_{-\tau,-\textbf{k},\downarrow}^\dagger,-c_{-\tau,-\textbf{k},\uparrow}^\dagger]^T$,
where $c_{\tau,\mathbf{k},s}$ annihilates an electron with valley index $\tau$, momentum $\mathbf{k}$, and spin $s$. We treat the gap $\Delta$ as a phenomenological order parameter without committing to a specific microscopic pairing mechanism. 

The first is an intervalley $s$-wave state described by
\begin{align}\label{eq:H_BdG_swave}
    h_{\rm BdG,\tau} = h_{0,\tau}\sigma_z + hs_z + \Delta\sigma_+ + \Delta^*\sigma_-,
\end{align}
where $h$ denotes the Zeeman splitting from a ferromagnetic substrate or external out-of-plane field, and $\sigma_{\pm}=\tfrac{1}{2}(\sigma_x\pm i\sigma_y)$.  
The second case is an intravalley chiral $p$-wave pair-density-wave state relevant to monolayer TMDs \cite{hsu_2017_topological} and rhombohedral graphene \cite{han_2024_signatures}.  
In the Ising limit ($\lambda_{so}=0$), the BdG Hamiltonian reads
\begin{align}\label{eq:H_BdG_pwave}
    h_{\rm BdG,\tau} = (\xi_{k}-\beta_{\rm so}+\tau h)\sigma_z + \Delta_{\tau}\sigma_+ + \Delta_{\tau}^*\sigma_-,
\end{align}
where the order parameter at each valley $\tau$ takes the form
\begin{align}\label{eq:chiral_Delta}
    \Delta_{\tau} =&\ \left\lbrace\begin{array}{cc}
       \Delta_0(k_x + ik_y),\ {\rm (same\ chirality)},  \\
       \Delta_0(k_x + i\tau k_y),\ {\rm (opposite\ chirality)}. 
    \end{array}\right.
\end{align} 
The same-chirality gap spontaneously breaks time-reversal symmetry (TRS) and exhibits a finite Chern number.

\begin{figure}[t!]
\includegraphics[width=0.5\textwidth]{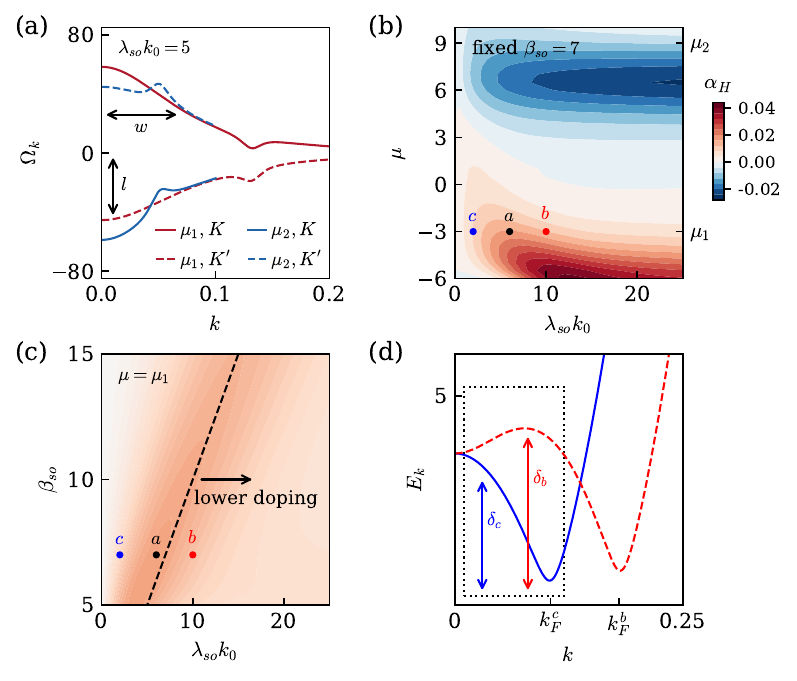}
\caption{Properties of intervalley $s$-wave pairing. (a) Berry curvature $\Omega_k$ for the $K$ (solid) and $K'$ (dashed) valleys at chemical potentials $\mu_1 = 4 - \beta_{\rm so}$ (red) and $\mu_2 = 2 + \beta_{\rm so}$ (blue) with the same $t$ and $\Delta$. See the corresponding doping regimes in Fig.~\ref{fig1}(b). The Nernst signal $\alpha_H(\lambda_{\rm so}k_0,\mu)$ 
as a function of Rashba strength $\lambda_{\rm so}k_0$, and the (b) chemical potential $\mu$ and (c) Ising SOC strength $\beta_{\rm so}$ (shared colorbar). Here, we set $k_0=\lambda_{\rm so}/2t$, $\beta_{\rm so}=7$, $t=500$, $T=1$, $\Delta=1$ and $h=0.5$. The dashed line in (c) marks the condition that maximizes $\alpha_H$, where Rashba and Ising SOC strengths are comparable. This line shifts to the right as doping $\mu$ lowers \cite{SM}. Representative points $a$, $b$, $c$ sit near and away from the line. (d) The lower BdG band dispersion $E_k$ at valley $K$ for representative points $b$ (red dashed) and $c$ (blue solid), where $k^{b/c}_F$ and $\delta_{b/c}$ are their Fermi momenta and relevant excitation gaps, respectively. The dotted box labels the main quasiparticle regime contributing to $\alpha_H$. Units: meV for energies, $a^{-1}$ ($a$ lattice constant) for $k$, $a^{-2}$ for $\Omega_k$, ${\rm meV}\cdot a^2$ for $t$, and $\alpha_0 = ek_B/(2\pi\hbar)$ for $\alpha_H$.\label{fig2}}
\end{figure}

The Berry curvature of this model, which governs the Nernst conductance $\alpha$ in Eq.~(\ref{eq:alpha_H}), is given by \cite{liao_2025_detecting}
\begin{align}\label{eq:BC}
    \Omega_{n{\bf k}} = -\frac{1}{2} \nabla_{\bf k} \rho_{n\bf k} \times \nabla_{\bf k} \chi_{\bf k} + \frac{1}{2} \nabla_{\bf k} s_{n\bf k} \times \nabla_{\bf k} \phi_{\bf k},
\end{align}
which originates from quasiparticle wavefunction twists in the particle–hole (first term) and spin (second term) spaces. Here, $s_{n\bf k} = \langle\psi_{n{\bf k}}|s_z|\psi_{n{\bf k}}\rangle$ is the state-resolved spin,  $\chi_{\bf k}=\arg(\Delta)$ is the gap phase, and $\phi_{\bf k}=\arctan(k_y/k_x)$ denotes the Rashba-induced spin winding. 
Rotational symmetry around momenta $K$ and $K'$ simplifies $\Omega_{n\mathbf{k}}$ to \cite{SM}
\begin{align}\label{eq:BC_rot}
    \Omega_{nk} = \left\lbrace\begin{array}{cc}
         \frac{1}{2k}\frac{ds_{nk}}{dk},\quad ({\rm intervalley\ \textit{s}-wave}), \\
         -\frac{1}{2k}\frac{d\rho_{nk}}{dk},\quad ({\rm intravalley\ \textit{p}-wave}). 
    \end{array}\right.
\end{align}
For the $s$-wave state, the $k$-independent gap with $\nabla_{\bf k}\chi=0$ makes the first term in Eq. (\ref{eq:BC}) vanish, whereas for the $p$-wave state with only Ising SOC, the second term vanishes due to the absence of Rashba coupling.

\textit{Intervalley s-wave paired state---} 
Since this paired state is non-chiral, an external Zeeman field or exchange coupling ($h \neq 0$) is required to break the TRS; without it, the two valleys contribute equally with opposite signs so that the total  $\alpha_H=0$. 

As shown in Fig.~\ref{fig2}(a), the Berry curvature magnitude maximizes at small momenta $k\sim0$ and its
sign flips between the two valleys by the Ising SOC, as well as between the lower and upper bands.  
Therefore, we predict a sign change in the Nernst signal $\alpha_H$ when changing the chemical potential from the lower to the upper band [see Fig.~\ref{fig2}(b)]. This qualitative signature can experimentally diagnose whether the observed Nernst effect is of Berry-curvature origin rather than from vortex motion. 

Next, we examine the conditions for optimizing the Nernst signal $\alpha_H$ of the lower band at chemical potential $\mu=\mu_1$ [see Fig.~\ref{fig1}(b)]. 
The Nernst response originates from thermally excited quasiparticles within an energy window near the Fermi level, where the entropy factor $-\partial g_{n}/\partial T$ in Eq.~(\ref{eq:alpha_H}) is sizable. 
Together with the fact that the Berry curvature peaks at small $k$, 
the signal magnitude is controlled by the following three quantities: (1) the width $w \approx |\beta_{\rm so}/2\lambda_{\rm so}|$ and (2) the height $l$  of the Berry curvature peak \footnote{The Berry curvature in the limit of small superconducting gap $\Delta\to 0$ is given by the standard form $\Omega_k = \frac{2\lambda_{\rm so}^2 h_{\tau}}{(h_{\tau}^2+4\lambda_{\rm so}^2 k^2)^{3/2}}$ (excluding the regime near the gap) \cite{Xiao2012}, where  $h_{\tau} = \tau\beta_{\rm so}+h$ the spin-orbit gap at $k=0$.}, and (3) the size of the quasiparticle gap $\delta$, which controls the number of thermally excited quasiparticles with large Berry curvatures. 

We find that the Nernst signal $\alpha_H$ is maximized when the Rashba SOC is comparable to the Ising SOC $\lambda_{\text{so}}k_0\sim\beta_{\text{so}}$ \footnote{$k_0=\lambda_{\rm so}/2t$ is a characteristic momentum in Rashba systems.}, labeled by the dashed line in Fig.~\ref{fig2}(c). This can be understood by comparing the representative points $a$, $b$, $c$  at a given $\beta_{\text{so}}$: At point $a$ where the two SOC strengths are  comparable, the width $w$ is comparable to the Fermi momentum $k_F^a$. The Berry curvature peak thus 
covers the momentum range $k\leq k_F^a$ where quasiparticles are predominantly excited. The quasiparticle Berry curvatures, as well as the Nernst signal $\alpha_H$, are therefore maximized.    
At point $b$ where Rashba becomes weaker than Ising SOC, since the peak width $w<k_F^b$ and the quasiparticle gap $\delta_b$ becomes larger at $k<k_F^b$, there are \textit{less} quasiparticles with larger Berry curvertures at a given temperature $T$. The Nernst signal $\alpha_H$ thus decreases. 
At point $c$ where Ising becomes weaker than Rashba SOC, although the peak width increases $w>k_F^c$ and the excitation gap $\delta_c$ decreases [see Fig. \ref{fig2}(d)], the peak height $l$ of Berry curvatures also decreases. Large-Berry-curvature quasiparticles thus become fewer from that at point $a$, and so does the Nernst signal $\alpha_H$. 

\textit{Intravalley chiral p-wave paired state---} 
When the electrons form pairs within a valley in a system with Ising SOC, the pairing gap is dictated to be equal-spin triplet and thus odd-parity in momentum $k$. Energetically, we expect that a nodeless chiral $p$-wave is favored over a nodal $p$-wave gap. Moreover, in this two-valley model with Ising SOC, such intravalley pairs also form a spin-triplet pair density wave since the pairs at valley $K$ and $K'$ each carries momentum $2K$ and $-2K$, respectively  \cite{hsu_2017_topological}. The chiral $p$-wave finite-momenta pairs at the two valleys can independently carry a Chern number $\pm1$.

In contrast to intervalley pairs, the quasiparticle Berry curvature $\Omega_{n\mathbf{k}}$ of intravalley pairs is nonzero even in the absence of an external magnetic field. 
This originates from the  phase $\chi_{\bf k}$ of the gap [see Eq.~(\ref{eq:chiral_Delta})] that winds in $k$ even in the absence of Rashba SOC. 
The identical or opposite Berry curvatures at the two valleys will thus lead to a \textit{spontaneous Nernst effect} and a \textit{spontaneous spin Nernst effect} due to the identical and opposite Chern numbers, respectively. 
Experimentally, we propose that the measurement of Nernst signals in charge and spin channels in the absence of external magnetic fields or exchange couplings is a qualitative smoking-gun evidence for chiral superconductivity.

We now discuss guiding principles for maximizing the Nernst signal for intravalley pairs. Unlike the intervalley pairs in Fig.~\ref{fig2}(a), the quasiparticle Berry curvature of intravalley pairs peaks at momentum $k=k_F$ 
due to the twist of the wavefunction in particle–hole space [see Fig.~\ref{fig3}(a)].
Counter-intuitively, although the quasiparticles excited at $k_F$ carry the largest magnitude of Berry curvature, they \textit{barely} contribute to the Nernst signal $\alpha_H$ in Eq. \eqref{eq:alpha_H}. 
This is because 
quasiparticles at $k_F$ consist of equal weights of particles and holes, whose contributions to $\alpha_H$ cancel out due to the vanishing quasiparticle charge $\rho_{n\mathbf{k}}$. 
Therefore, we propose to maximize $\alpha_H$ by operating at \textit{a finite temperature $T$ closer to $T_c$} to  
excite higher-energy quasiparticles with particle-hole asymmetry [see the box in Fig.~\ref{fig3}(b)] without destroying superconductivity. In Fig.~\ref{fig3}(c), we numerically show that the Nernst signal $\alpha_H$ for the chiral $p$-wave intravalley pairs peaks at a temperature $T^*\sim 0.8 T_c$. This behavior reflects the interplay between quasiparticle excitation and TRS breaking: below $T^*$, the number of particle-hole asymmetric quasiparticles with large Berry curvature is limited, while above $T^*$ superconductivity is suppressed and TRS is restored, leading to a reduction of $\alpha_H$. In contrast, for the $s$-wave case, $\alpha_H$ increases monotonically with temperature up to $T_c$ \cite{SM}, since TRS is broken by an external field that persists across the superconducting transition. Notably, $\alpha_H$ for the chiral TSC can exceed that of $s$-wave pairs by an order of magnitude (see Table~\ref{tab:summary}), where TRS is only weakly broken by the Pauli- or orbital-limited field \footnote{The Nernst response for the $s$-wave state can in principle exceed that of the chiral $p$-wave state at large Zeeman fields in the absence of orbital effects, but this regime is typically inaccessible once orbital pair breaking is included. In Fig.~\ref{fig2}, we choose $h=0.5$ for the $s$-wave case, which is below the Pauli limit to ensure the stability of the superconducting state, although it may still exceed the orbital-limited critical field in realistic systems. See Supplementary Materials for more details.}.

\begin{figure}[t!]
\includegraphics[width=.45\textwidth]{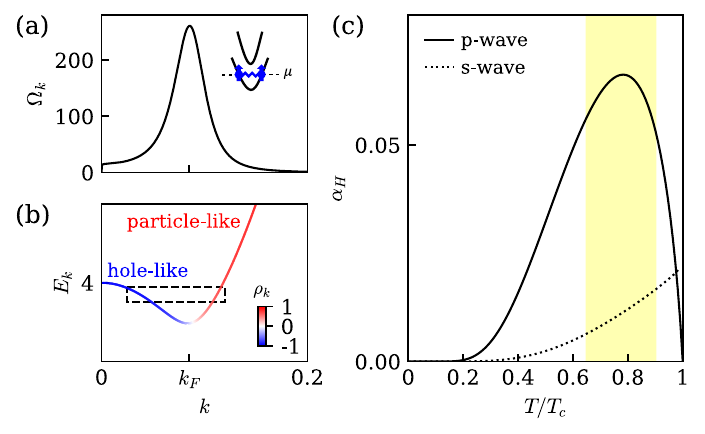}
\caption{Properties of intravalley chiral $p$-wave pairing. (a) Berry curvature and (b) the corresponding lower BdG band at the $K$ valley, with color indicating the quasiparticle charge $\rho_k$. Here, $t=500$, $\mu = 4 - |\beta_{\rm so}|$, and $\Delta_0 k_F = 2$, with Fermi momentum $k_F \approx 0.85$. (c) Nernst signal $\alpha_H(T)$ at fixed Ising SOC with no Rashba or external field ($\lambda_{\rm so}=h=0$). The solid (dashed) curve denotes chiral $p$-wave ($s$-wave) pairing. The $s$-wave parameters are chosen as in point $a$ of Fig.~2(b), so that both cases have the same doping and critical temperature $T_c \approx 1.11$ at $h=0$. For the $p$-wave case, we use the BCS form $\Delta_0(T) k_F \approx 1.76 k_B T_c \tanh(1.74\sqrt{T_c/T - 1})$~\cite{tinkham2004introduction}, while for the $s$-wave case the temperature dependence of the gap is obtained by solving the gap equation \cite{SM}. The shaded region highlights the temperature range maximizing $\alpha_H$. Units are the same as in Fig.~\ref{fig2}.\label{fig3}}
\end{figure}

\newcolumntype{C}{>{\centering\arraybackslash}X}

\begin{table}[t!]
\small
\centering
\caption{Summary of the origin and hotspot of Berry curvature $\Omega_{\bf k}$, time-reversal-breaking mechanism that leads to Nernst conductance $\alpha_H\neq 0$, typical $\alpha_H$ magnitude in unit of $\alpha_0\approx 3.3\,\mathrm{nA/K}$, the loop current density $J\propto L^{-1}\Delta T$, and magnetic field $B_{\rm th} \propto R^{-1}\Delta T$ at the ring center for a ring radius $R \approx 60\,\mathrm{nm}$ under a temperature gradient $\Delta T \approx 1\,\mathrm{K}$ across a radial distance $L = 100\mathrm{nm}$.}
{
\renewcommand{\arraystretch}{1.25} 
\begin{tabularx}{\linewidth}{@{}p{2.2cm}CC@{}}
\hline\hline  
 & $s$-wave & chiral $p$-wave \\
\hline
$\Omega_{\bf k}$ origin & spin texture from spin-orbit interaction & pseudospin texture from chiral pairing \\
\hline
$\Omega_{\bf k}$ hotspot & $k\approx0$ & $k\approx k_F$ \\
\hline
time-reversal-breaking origin & external magnetic field & spontaneous chiral pairing \\
\hline
Nernst signal $\alpha_H$  & $O(10^{-2})$ & $O(10^{-1})$ \\
\hline
$J$ magnitude & $10^{-3}\mathrm{A/m}$ & $10^{-2}\mathrm{A/m}$ \\
\hline
$B_{\rm th}$ magnitude  & $\mathrm{nT}$ & $10\,\mathrm{nT}$ \\
\hline\hline  
\end{tabularx}
}
\label{tab:summary}
\end{table}

\textit{Experimental setup and estimation---\label{sec:discussion}} First, we propose an experimental setup to measure the predicted Nernst effect that avoids suppression due to screening. When the sample is in a bar geometry, the transverse current can cause charge accumulation and motion on the edges, which are expected to be screened by backflows in superconductors, leading to vanishing Nernst effect. To circumvent this  screening effect, we propose to measure the Nernst conductance $\alpha_H$ in a ring geometry fabricated by a quasi-2D strongly type-II superconductor [see Fig.~\ref{fig1}(a)]. In this setup, the transverse charge current activated by an  applied temperature gradient in the radial direction will flow along the ring, generating a magnetic flux $\Phi_{\rm th}$ threading through the ring center. Note that the flux $\Phi_{\rm th}$ is \textit{non-quantized} even in the case of chiral topological superconductors with a finite Chern number. This is because in contrast to the thermal conductivity, which is well-known to be quantized in chiral superconductors owing to the Majorana edge modes responsible for heat transport \cite{read_2000_paired}, Nernst conductance $\alpha_H$ is originated from thermally activated non-Majorana quasiparticles near the superconducting gap, manifested by the distribution factor $dg/dT$ in Eq.~(\ref{eq:alpha_H}). Nernst effect is thus a consequence of the band geometry rather than the band topology characterized by the Chern numbers. 

Using the setup in Fig.~\ref{fig1}(a), 
a chiral $p$-wave TSC candidate can be experimentally identified by 
a spontaneous appearance of flux $\Phi=\Phi_{\rm th}$ in the absence of an external magnetic field. In contrast, 
for a ring fabricated by a spin-orbit coupled $s$-wave superconductor, 
such a Nernst-induced flux $\Phi_{\rm th}$ can only be triggered by an out-of-plane magnetic field. Specifically, we expect a larger-than-expected flux at the ring center $\Phi=n\Phi_0+\Phi_{\rm th}$, consisting of field-induced $n$ flux quanta $\Phi_0$ in additional to the Nernst contribution.

To experimentally test our prediction, the ring in Fig.~\ref{fig1}(a) can be fabricated using quasi-2D superconductors of interest, where the radius is smaller than the penetration depth. In this regime, vortex contributions to the Nernst effect are naturally suppressed. Away from this ideal limit, cases such as thicker samples exhibiting stronger magnetic-field screening require more careful treatment since the penetration depth and its temperature dependence can influence the thermally induced magnetic flux, as observed in the quasiparticle Seebeck effect in a bimetallic ring geometry \cite{shelly_2016_resolving}. A microscopic theory that integrates the screening effect into the semiclassical framework is an interesting future work.

The competition between intrinsic and extrinsic contributions to the superconducting Nernst effect remains an important open question. A systematic theory of extrinsic mechanisms, analogous to side-jump and skew-scattering contributions in anomalous Hall transport, has not yet been fully developed. Moreover, unlike electrical transport, thermal transport coefficients depend not only on the relaxation time but also explicitly on temperature through the quasiparticle distribution and excitation spectrum, making a simple scaling-based separation of different contributions unreliable.

Finally, we estimate the magnitudes of the spontaneous and field-activated Nernst signals in the $p$- and $s$-wave cases, respectively. Consider first an $s$-wave superconductor with a thermoelectric conductance $\alpha_H = 0.03\alpha_0 \approx 0.1\,\mathrm{nA/K}$ \footnote{We adopt a representative value from Fig.~\ref{fig2}(b) for the $s$-wave case.}. For a temperature difference $\Delta T = 1\,\mathrm{K}$ across a radial distance $L = 100\,\mathrm{nm}$ (set by the ring geometry or the London penetration depth), the resulting loop current $I = \alpha_H \Delta T$, corresponding to a current density $J = I/L \approx 10^{-3}\,\mathrm{A/m}$. In a ring of inner radius $R = 60\,\mathrm{nm}$, the corresponding magnetic field at the center is $B_{\rm th} = \mu_0 I / (2R) \approx 1\,\mathrm{nT}$ ($\mu_0$ the vacuum permeability), and it increases toward the ring boundary. By contrast, for a chiral $p$-wave superconductor with $\alpha_H = 0.3\alpha_0 \approx 1\,\mathrm{nA/K}$ \footnote{We adopt a value an order of magnitude larger than the $s$-wave case, consistent with Fig.~\ref{fig3}(c). The choice $\alpha_H = 0.3$ slightly exceeds the peak value in Fig.~\ref{fig3}(c), reflecting that the Nernst response can be further enhanced in materials with lower doping, higher $T_c$, or in thicker samples supporting more subbands.}, the same $\Delta T$ and geometry give $J \approx 10^{-2}\,\mathrm{A/m}$ and $B_{\rm th} \approx 10\,\mathrm{nT}$. Smaller geometries ($L$ or $R$) yield stronger local signals but require higher spatial resolution, accessible with nanoscale magnetometry using single nitrogen-vacancy (NV) centers \cite{Taylor2008, Maletinsky2012}, whereas larger rings generate greater thermal flux $\Phi_\mathrm{th}$ that can be detected with high-sensitivity magnetometers such as ensemble NV centers \cite{barry_2020_sensitivity} or SQUIDs \cite{Clarke2004}.
The Berry-curvature origin, time-reversal-breaking mechanisms, and expected magnitudes of the Nernst-induced current and magnetic field in the ring geometry are summarized in Table~\ref{tab:summary}.

\textit{Summary---}  We show that Nernst effect in superconductors is an understudied transport property that can diagnose the pairing symmetry. 
Chiral topological superconductors can be identified by a spontaneous Nernst flux $\Phi_{\rm th}$ in our proposed setup in Fig.~\ref{fig1}(a), in the absence of external magnetic field. In contrast to thermal conductivity, the existence of the flux $\Phi_{\rm th}$, rather than its quantization, is enough to diagnose a chiral superconductor. 
For non-chiral superconductors, we show that Nernst effect can be activated by an external magnetic field
even with an $s$-wave gap, as long as the material has both Ising and Rashba spin-orbit couplings. Sizable Nernst signals, namely the flux $\Phi_{\rm th}$, can be achieved by choosing materials systems with comparable Ising and Rashba spin-orbit couplings along with controlling the chemical potential to lie in the spin-orbit gap. Finally, we expect nodal superconductors to exhibit an enhanced Nernst signal due to the abundant  quasiparticles, making this an interesting direction for future study.   

\textit{Acknowledgments---} 
T.-C.H. acknowledges support from Department of Energy Basic Energy Science Award No. DE-SC0024291 for tuning Berry curvatures in superconductors with spin-orbit-couplings as well as activating Nernst effect with external fields. Y.-T.H. acknowledges support from Department of Energy Basic Energy Science Award No. DE-SC0026108 for identifying experimental signatures of topological superconductors. C.X. acknowledges support from the start-up funding from Fudan University and National Natural Science Foundation of China (12574114).

\textit{Data availability---} There are no publicly available research data or software supporting this manuscript. Requests for further information or data should be sent to the authors. \cite{Oh2024,Xiao2012}

\appendix
\begin{widetext}
\title{Supplemental Material for Intrinsic Nernst Effect from Berry Curvature in Superconductors}

\author{Tzu-Chi~Hsieh}\affiliation{Department of Physics, University of Notre Dame, Notre Dame, Indiana 46556, USA}
\author{Cong Xiao}\affiliation{Interdisciplinary Center for Theoretical Physics and Information Sciences (ICTPIS), Fudan University, Shanghai 200433, China}
\author{Yi-Ting~Hsu}\affiliation{Department of Physics, University of Notre Dame, Notre Dame, Indiana 46556, USA}
\date{\today}

\maketitle

\section{Derivation of Eq.~(2) in the Main Text}

For completeness, we outline here the main steps of the derivation of the Nernst conductance $\alpha_H$, which closely follow Refs.~\cite{DongPRL2020,XiaoPRB2021}. The detailed calculations can be found therein. Starting from Eq.~(\ref{eq:qp_continuity}) in the main text,
\begin{align}\label{eq:continuity_app}
    \nabla_{\mathbf{r}} \!\cdot \mathbf{j}_{\mathrm{qp}}
    = \mathcal{B}_{p}
    = -\,\nabla_{\mathbf{r}} \!\cdot \mathbf{j}_{s},
\end{align}
the Nernst conductance is obtained by evaluating the linear response of the total transport current
$\mathbf{j} = \mathbf{j}_{\mathrm{qp}} + \mathbf{j}_{s}$
to a temperature gradient. This requires a first-order gradient expansion of the quasiparticle
current $\mathbf{j}_{\mathrm{qp}}$ and a second-order expansion of the source term $\mathcal{B}_{p}$.

To this end, we employ a variational approach to compute the expectation value of the relevant
quantities over a wavepacket state
$\lvert \Phi(\mathbf{k}_{c}, \mathbf{r}_{c}, t) \rangle$
centered at the phase-space coordinate $(\mathbf{k}_{c}, \mathbf{r}_{c})$ and time $t$:
\begin{equation}
\left. \frac{\delta S}{\delta h} \right|_{P} (\mathbf{r}, t)
= -\!
\left.
\frac{
    \langle \Phi(\mathbf{k}_{c}, \mathbf{r}_{c}, t) \vert
    \hat{\theta} \,\delta(\hat{\mathbf{r}} - \mathbf{r}) \vert
    \Phi(\mathbf{k}_{c}, \mathbf{r}_{c}, t) \rangle
}{
    \langle \Phi(\mathbf{k}_{c}, \mathbf{r}_{c}, t) \vert
    \Phi(\mathbf{k}_{c}, \mathbf{r}_{c}, t) \rangle
}
\right|_{P}.
\end{equation}

Here, $h(\mathbf{r}, t)$ is an auxiliary field that couples to the physical observable
$\hat{\theta}$ via the term $\hat{\theta}\cdot h(\mathbf{r}, t)$.
The action is defined as
$S = \int d\tau\, \mathcal{L}$,
where the wavepacket Lagrangian is $\mathcal{L} = \langle \Phi | i\partial_{\tau} - \hat{H} | \Phi \rangle / \langle \Phi | \Phi \rangle$. The operator $\hat{H}$ represents the local Hamiltonian experienced by the wavepacket, 
expanded around $\mathbf{r}_{c}$ up to second order in spatial gradients. The subscript $\lvert_{P}$ denotes evaluation for physical wavepackets $\lvert\Phi(\mathbf{k}_{c}, \mathbf{r}_{c}, t)\rangle$ satisfying the Schr\"odinger equation,
and thus, semiclassically, the equations of motion for the wavepacket center
$(\mathbf{k}_{c}, \mathbf{r}_{c})$.
Throughout this supplementary material, we put the ``hat" symbol to indicate quantum operators.

The above expression represents the contribution of a single electron to the $\theta$ density. 
By summing over all electronic contributions via the phase-space integration 
$\int [d\mathbf{k}_c]\, d\mathbf{r}_c\, D f_{\mathrm{tot}}$, 
where $D(\mathbf{k}_{c}, \mathbf{r}_{c})$ is the Berry-curvature--corrected phase-space measure, 
$f_{\mathrm{tot}}(\mathbf{k}_{c}, \mathbf{r}_{c})$ is the phase-space occupation function, 
and $[d\mathbf{k}_c]$ denotes 
$\sum_{n} d\mathbf{k}_c / (2\pi)^{d}$ 
with $d$ being the spatial dimensionality 
(we adopt the convention $\hbar = 1$ here), 
the local density of $\theta$ ig given by
\begin{align}
    \theta_s^{\mathrm{loc}}(\mathbf{r}, t)
    = - \int [d\mathbf{k}_c]\, d\mathbf{r}_c\, D f_{\mathrm{tot}}
    \left.\frac{\delta S}{\delta h_s}\right|_{P},
\end{align}
where the subscript $s$ labels different observables. In the following discussion, the notation $\lvert_{P}$ will be omitted for brevity.

After a tedious but systematic evaluation of the wavepacket Lagrangian and the field variation $\delta S/\delta h$, the local density is found to be
\begin{align}
\theta_s^{\mathrm{loc}}(\mathbf{r}, t)
&= \int D f_{\mathrm{tot}}
\left( 
    \frac{\partial \tilde{\epsilon}}{\partial h_s}
    - \tilde{\Omega}_s^{h \mathcal{T}}
\right)
- \partial_i \int D f_{\mathrm{tot}}
\left[
    \tilde{d}_{is}^{\theta}
    - \dot{k}_j \frac{\partial a_j^{k}}{\partial (\partial_i h_s)}
    - \left( r_i \partial_l h_j + \partial_t h_j \right)
      \frac{\partial a_j^{h}}{\partial (\partial_i h_s)}
\right]
+ \partial_{ij} \int D f_{\mathrm{tot}}q_{ijs}^{\theta},
\end{align}
where $i,j$ denote spatial Cartesian coordinates. Here, $\tilde{\Omega}_s^{h \mathcal{T}} 
  = \tilde{\Omega}_{si}^{hk} k_i
  + \tilde{\Omega}_{si}^{hr} r_i
  + \tilde{\Omega}_s^{ht}$
is the Berry curvature associated with the total time derivative, with $\tilde{\Omega}_{si}^{hk}$ representing the Berry curvature in $h_s-k_i$ parameter space as an example. $\tilde{\epsilon}$ is the wavepacket energy, $\tilde{d}_{is}^{\theta}$ is the dipole moment, $q_{ijs}^{\theta}$ is the quadrupole moment, and $a_j^{k(h)}$ is the positional shift, which can be viewed as the linear-order correction to the Berry connection in $k(h)$ space. The ``tilde" emphasizes the inclusion of the correction up to the second-order gradient expansion, and the explicit forms of those quantities can be found in the supplementary materials of Ref.~\cite{XiaoPRB2021}. Hereafter, the integral $\int$ will refer to integration over momentum space for notational simplicity.

The distribution function of semiclassical Bloch electrons is given by
\begin{align}
    f_{\mathrm{tot}} = f + \delta f,
    \label{eq:ftot}
\end{align}
where $f = f(\tilde{\epsilon})$ is the equilibrium Fermi-Dirac distribution, and 
$\delta f$ denotes the out-of-equilibrium correction arising from scattering processes, 
as determined by the semiclassical Boltzmann equation. Consequently, $\theta_s^{\mathrm{loc}}(\mathbf{r}, t)$ can be decomposed into 
intrinsic and extrinsic contributions, associated respectively with $f$ and $\delta f$. In the following, we focus on the intrinsic part of the local density and comment on the extrinsic contribution at the end.

Now consider the local equilibrium distribution $f$, which incorporates 
slowly varying temperature $T$ and chemical potential $\mu$ in real space. 
After tedious calculations and setting the auxiliary field to zero, this leads to the following expression for the local density:
\begin{align}\label{eq:theta_loc}
\theta_s^{\mathrm{loc,in}}
&= \int f^0 \theta_s 
- \int \Omega_{sj}^{hk} F_j(\epsilon)
- \partial_i 
\left\{
    \int \left( f^0 d_{is}^\theta + g^0 \Omega_{is}^{kh} \right)
    + \int \chi^{ils}F_j(\epsilon)
    - \partial_j \int 
    \left[ f^0 q_{ijs}^\theta + g^0 \chi^{ils} \right]
\right\}
\end{align}
where $f^0 = f(\epsilon)$, the grand potential density $g^0 = k_B T\ln(1-f^0)$, $\theta_s({\bf k})$ is the average over the eigenstate of the BdG Hamiltonian, the $\theta_s$-dipole polarizability of a bulk semiclassical electron,
\begin{align}
    \chi^{ils}= -\left.\frac{\partial a_j^k}{\partial (\partial_i h_s)}\right|_{h_s=0},
\end{align}
and
\begin{equation}
F_j(\epsilon)
= f^0 (eE_j - \partial_j \mu)
- \frac{1}{T} \left[ (\epsilon - \mu) f^0 - g^0 \right] \partial_j T,
\end{equation}
with ${\bf E}$ the electric field. Note that the local density is now expressed in terms of the leading-order Berry curvature, wavepacket energy, dipole, and quadrupole moments, evaluated at the energy band $n$ of interest (for simplicity, we neglect the subscript $n$, which appears in the main text).

Next, by applying the semiclassical gradient-expansion formula [Eq.~(\ref{eq:theta_loc})] to the quantities appearing in the continuity equation [Eq.~(\ref{eq:continuity_app})], 
and retaining terms up to first order in spatial derivatives for $\hat{\mathbf{j}}_{\mathrm{qp}} = \frac{1}{2}\{\hat{{\bf v}},\hat{\rho}\}$ 
and up to second order for the source term $\hat{\mathcal{B}}_{p} = i[\hat{H_p}, \hat{\rho}]$, we can obtain
\begin{align}
    \mathbf{j}^\mathrm{loc} =  \mathbf{j}^\mathbf{M} + \mathbf{j},
\end{align}
where $\mathbf{j}^\mathbf{M} = \nabla\times\mathbf{M}$ is the magnetization current ($\mathbf{M}$ the orbital magnetization) and $\mathbf{j}$ is the transport current arising from the inhomogeneous temperature and chemical potential, the $F_j(\epsilon)$ terms in Eq.~(\ref{eq:theta_loc}). The quasiparticle and Cooper pair contributions of the latter are respectively given by
\begin{align}
    j_{\rm qp,i} = \int F_{l}\Omega_n^{li\rho},\quad j_{\rm s,i} = \int F_{l}\chi_n^{il\mathcal{B}},
\end{align}
where the $\rho$-Berry curvature and the $\mathcal{B}$-dipole polarizability are given by
\begin{align}
    \Omega_n^{li\rho} = -2\,\mathrm{Im}\sum_{n_1 \ne n}\frac{
    (v_l)_{n n_1}\tfrac{1}{2} \{ \mathbf{v}_i, s_j \}_{n_1 n}}{
    (\epsilon_n - \epsilon_{n_1})^2},
    \quad\chi_n^{il\mathcal{B}} = - \Omega^{li\rho} - (\Omega_{il})_n (s_j)_n + \frac{\partial}{\partial k_l} d_{ij}^{s}.
\end{align}
Combine both contributions gives the total transport current
\begin{align}
j_i
&= \sigma_{\mathrm{in}}^{ils} (E_l - \partial_l \mu / e)
    - \alpha_{\mathrm{in}}^{ils} \, \partial_l T,
\end{align}
with intrinsic conductance $\sigma_{\mathrm{in}}^{ils}$ and Nernst coefficient
\begin{align}
\alpha_{\mathrm{in}}^{ils}
= - \int \frac{\partial g_n}{\partial T}
    \left( \Omega_{n}^{li} s_n + \partial_{k_l} d_{n}^{is} \right).
\end{align}
Finally, it is found that in the presence of an arbitrary weak disorder potential, the dipole term is canceled out by the extrinsic contribution obtained by the relevant Boltzmann equation in the lowest
Born order \cite{XiaoPRB2021}. This results in the Eq.~(\ref{eq:alpha_H}) of the main text.

\section{Calculation Details of Berry curvature $\Omega_{n{\bf k}}$}

The Berry curvature (\ref{eq:BC}) of the main text consists of separable contributions from the twist of wavefunctions in particle-hole and spin spaces in the presence of an out-of-plane magnetic field. This stems from the property of the BdG Hamiltonian $h_{\rm BdG,\tau}$ that takes real form in a rotated frame around the z-axis $\tilde{h}_{\rm BdG,\tau}=U_{\tau}h_{\rm BdG,\tau}U_{\tau}^\dagger$. For intervalley $s$-wave pairing with BdG Hamiltonian (\ref{eq:H_BdG_swave}) of the main text, the unitary operator $U = e^{-i\sigma_z\chi_{\bf k}/2}e^{-is_z(\phi_{\bf k}-\pi/2)/2}$ is valley independent (with subscript $\tau$ dropped) and the real BdG Hamiltonian
\begin{align}
    \tilde{h}_{\rm BdG,\tau} =&\ \left(\begin{array}{cccc}
        \tau\beta_{\rm so} + h + \xi_k & \lambda_{\rm so}k & |\Delta| & 0 \\
        \lambda_{\rm so}k & -\tau\beta_{\rm so} - h + \xi_k & 0 & |\Delta| \\
         |\Delta| & 0 & -\tau\beta_{\rm so} + h - \xi_k & -\lambda_{\rm so}k \\
         0 & |\Delta| & -\lambda_{\rm so}k & \tau\beta_{\rm so} - h - \xi_k
    \end{array}\right).
\end{align}
This determines the form of the eigenfunction, given by
\begin{align}\label{eq:wavevector_swave}
    |\psi_{n\tau}\rangle = U^\dagger|\tilde{\psi}_{n\tau}\rangle = \left(\begin{array}{c}
         \tilde{u}_{n\tau\uparrow}e^{i(\chi-\phi+\pi/2)/2}  \\
         \tilde{u}_{n\tau\downarrow}e^{i(\chi+\phi-\pi/2)/2}  \\
         \tilde{v}_{n\bar{\tau}\downarrow}e^{-i(\chi+\phi-\pi/2)/2}  \\
         \tilde{v}_{n\bar{\tau}\uparrow}e^{-i(\chi-\phi+\pi/2)/2}
    \end{array}\right),
\end{align}
where $\bar{\tau} \equiv -\tau$, $|\tilde{\psi}_{n\tau}\rangle = \left(\begin{array}{cccc}
    \tilde{u}_{n\tau\uparrow} & \tilde{u}_{n\tau\downarrow} & \tilde{v}_{n\bar{\tau}\downarrow} & \tilde{v}_{n\bar{\tau}\uparrow}
\end{array}\right)^T$ is a purely real eigenvector of $\tilde{h}_{\rm BdG,\tau}$ labeled by band index $n$, and we have dropped the momentum argument ${\bf k}$ for the simplicity of the notation. With the phase structure of the wavevector (\ref{eq:wavevector_swave}), the Berry connection and Berry curvature are respectively given by
\begin{align}
    \mathcal{A}_{n\tau{\bf k}} =&\ i\langle\psi_{n\tau{\bf k}}|\nabla_{\bf k}|\psi_{n\tau{\bf k}}\rangle = - \frac{1}{2}\rho_{n\tau{\bf k}}\nabla_{\bf k}\chi_{\bf k} + \frac{1}{2}s_{n\tau{\bf k}}\nabla_{\bf k}\phi_{\bf k},
    \\
    \Omega_{n\tau{\bf k}} =&\ \nabla_{\bf k}\times \mathcal{A}_{n\tau{\bf k}} = -\frac{1}{2} \nabla_{\bf k} \rho_{n\tau{\bf k}} \times \nabla_{\bf k} \chi_{\bf k} + \frac{1}{2} \nabla_{\bf k} s_{n\tau{\bf k}} \times \nabla_{\bf k} \phi_{\bf k},
\end{align}
as derived in Ref.~\cite{liao_2025_detecting}. As a side note, crystal field and higher-order terms in ${\bf k}$ can lead to correction in $\phi_{\bf k}$ but not the structure of the wavefunction (\ref{eq:wavevector_swave}) and thereby Berry curvature (\ref{eq:BC}) of the main text.

For intravally chiral $p$-wave pairing, we consider the limit of vanishing Rashba effect $\lambda_{so}=0$ and focus on chemical potential lying within the lower energy bands, described by BdG Hamiltonian (\ref{eq:H_BdG_pwave}) of the main text. In this case, the unitary operator $U_{\tau} = e^{-i\sigma_z\chi_{\tau}/2}$, the real form BdG Hamiltonian
\begin{align}
    \tilde{h}_{\rm BdG,\tau} = \left(\begin{array}{cc}
        \xi_k - \beta_{so} - \tau h & |\Delta_0|k \\
        |\Delta_0|k & -\xi_k + \beta_{so} + \tau h 
    \end{array}\right),
\end{align}
and eigenvectors
\begin{align}
    |\psi_{K}\rangle = U_{K}^\dagger|\tilde{\psi}_{K}\rangle = \left(\begin{array}{c}
         \tilde{u}_{K\downarrow}e^{i\chi_K/2}  \\
         \tilde{v}_{K\downarrow}e^{-i\chi_K/2}  
    \end{array}\right),\quad |\psi_{K'}\rangle = U_{K'}^\dagger|\tilde{\psi}_{K'}\rangle = \left(\begin{array}{c}
         \tilde{u}_{K'\uparrow}e^{i\chi_{K'}/2}  \\
         \tilde{v}_{K'\uparrow}e^{-i\chi_{K'}/2}  
    \end{array}\right),
\end{align}
where $|\tilde{\psi}_{\tau}\rangle = \left(\tilde{u}_{\tau}\ \tilde{v}_{\tau}\right)^T$ is a purely real eigenvector of $\tilde{h}_{\rm BdG,\tau}$, with the band index dropped as there is only one normal band here, $\chi_{\tau{\bf k}} = \phi_{\bf k} + \chi_0$ for same chirality and $\chi_{\tau{\bf k}} = \tau\phi_{\bf k} + \chi_0$ for same opposite chirality at the two valleys, respectively, and $\chi_0 = \arg(\Delta_0)$ a constant phase. Accordingly, the Berry curvature is given by
\begin{align}
    \Omega_{\tau{\bf k}} = -\frac{1}{2} \nabla_{\bf k} \rho_{\tau{\bf k}} \times \nabla_{\bf k} \chi_{\tau\bf k},
\end{align}

Finally, for the low-energy Hamiltonians (\ref{eq:H_BdG_swave}) and (\ref{eq:H_BdG_pwave}) of the main text considered here, the rotational symmetry ensures $\nabla_{\bf k}\rho_{n{\bf k}}=\partial_k\rho_{nk}\hat{k}$ and $\nabla_{\bf k}s_{n{\bf k}}=\partial_k s_{nk}\hat{k}$. Together with the momentum dependence of the phases $\chi_{\bf k}\sim\phi_{\bf k} = \arctan (k_y/k_x)$, the Berry curvature reduces to Eq.~(\ref{eq:BC_rot}) of the main text.

\section{Gap equation for s-wave intervalley pairing}

For s-swave intervalley pairing, We consider the BdG Hamiltonian introduced in Eq.~(\ref{eq:H_BdG_swave}) of the main text, which includes Ising and Rashba spin-orbit couplings as well as a Zeeman field (orbital effects are neglected).

To describe superconductivity, we introduce an intervalley spin-singlet pairing interaction
\begin{equation}
H_{\mathrm{int}}=-U\sum_{\mathbf{k},\mathbf{k}'} B_{\mathbf{k}}^\dagger B_{\mathbf{k}'},
\end{equation}
where $U>0$ and
\begin{equation}
B_{\mathbf{k}}=
c_{-\!, -\mathbf{k},\downarrow}c_{+\!, \mathbf{k},\uparrow}
-
c_{-\!, -\mathbf{k},\uparrow}c_{+\!, \mathbf{k},\downarrow}.
\end{equation}
Under mean-field decoupling, this leads to
\begin{equation}
H_{\mathrm{pair}}=
\sum_{\mathbf{k}}
\left[
\Delta B_{\mathbf{k}}^\dagger+\Delta^* B_{\mathbf{k}}
\right]
+\frac{|\Delta|^2}{U},
\end{equation}
with the self-consistency condition
\begin{equation}
\Delta=-U\sum_{\mathbf{k}}\langle B_{\mathbf{k}}\rangle.
\end{equation}

In the numerical implementation, we work within a single valley sector and choose $\tau=+1$. The BdG Hamiltonian is evaluated in the Nambu basis
\begin{equation}
\Psi(\mathbf{k})=
\bigl[
c_{+,\mathbf{k},\uparrow},
c_{+,\mathbf{k},\downarrow},
c^\dagger_{-,-\mathbf{k},\downarrow},
- c^\dagger_{-,-\mathbf{k},\uparrow}
\bigr]^T,
\end{equation}
consistent with Eq.~(\ref{eq:H_BdG_swave}) of the main text. For each momentum $k$, we diagonalize the BdG Hamiltonian,
\begin{equation}
h_{\mathrm{BdG}}(k)\,\psi_n(k)=E_n(k)\,\psi_n(k),
\end{equation}
and evaluate the anomalous expectation value using the BdG eigenvectors. Here $\psi_n(k) = (\psi_{n1}, \psi_{n2}, \psi_{n3}, \psi_{n4})^T$ denotes the eigenvector of the BdG Hamiltonian expressed in the Nambu basis defined above. In this basis, the pairing amplitude takes the form
\begin{equation}
\langle B_{\mathbf{k}}\rangle
=
\sum_n F_n(k)\, f(E_n(k),T),
\end{equation}
with
\begin{equation}
F_n(k)=\psi_{n1}(k)\psi_{n3}^*(k)+\psi_{n2}(k)\psi_{n4}^*(k),
\end{equation}
and
\begin{equation}
f(E,T)=\frac{1}{e^{E/T}+1}.
\end{equation}
Substituting into the self-consistency condition leads to
\begin{equation}
\Delta = U\, J(\Delta,h,T),
\end{equation}
where
\begin{equation}
J(\Delta,h,T)
=
-\int\frac{k\,dk}{2\pi}
\sum_n F_n(k)\,f(E_n(k),T).
\end{equation}

To eliminate the bare interaction strength, we express the gap equation in terms of the reference gap $\Delta_0 \equiv \Delta(h=0,T=0)$ and define
\begin{equation}
\chi(\Delta,h,T)=\frac{J(\Delta,h,T)}{\Delta}.
\end{equation}
The self-consistency condition is then written in the renormalized form
\begin{equation}
\chi(\Delta,h,T)=\chi(\Delta_0,0,0).
\end{equation}
For fixed $(h,T)$, we numerically determine all positive solutions by locating the roots of
\begin{equation}\label{eq:Delta(h,T)}
\chi(\Delta,h,T)-\chi(\Delta_0,0,0)=0.
\end{equation}
The thermodynamically stable solution is selected by minimizing the free-energy difference
\begin{equation}
\Delta F(\Delta)
=
\int_0^\Delta dx\, 2x\left[\chi(\Delta_0,0,0)-\chi(x,h,T)\right].
\end{equation}
If no positive solution yields $\Delta F<0$, the equilibrium state is taken to be the normal state with $\Delta=0$.

\section{Supplementary plots for intervalley s-wave pairing}

\begin{figure*}[h]
\includegraphics[width=1\textwidth]{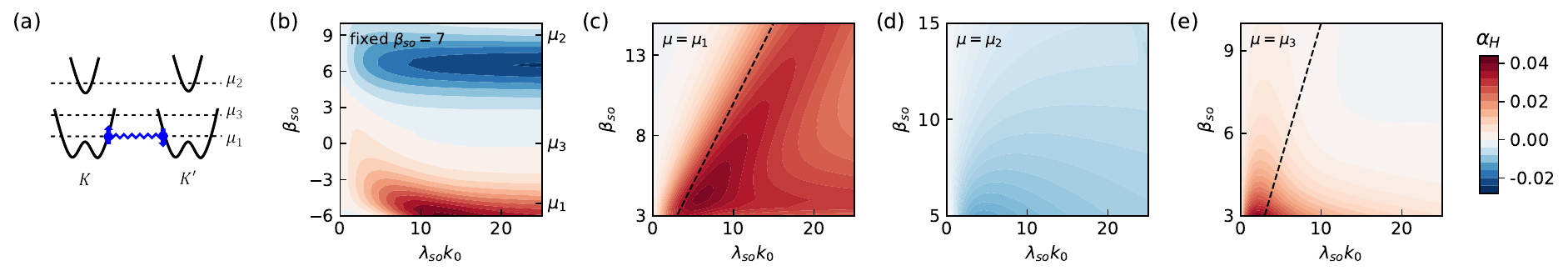}
\caption{(a) Intervalley $s$-wave pairing of the dominant low-energy bands in a type-II Ising superconductor. $K$ and $K'$ are time-reversal-related valleys, and $\mu$ is the chemical potential. The subscripts of $\mu$ label three distinct doping regimes. (b) $\alpha_H(\lambda_{\rm so}k_0,\mu)$ with $k_0=\lambda_{\rm so}/2t$, $\beta_{\rm so}=7$, $t=500$, $T=1$, $\Delta=1$ and $h=0.5$. (c)-(e) $\alpha_H(\lambda_{\rm so}k_0,\beta_{\rm so})$ with the same parameters as in (a), but with chemical potentials $\mu_1 = 2 - \beta_{\rm so}$, $\mu_2 = 2 + \beta_{\rm so}$, $\mu_3=0$, corresponding respectively to the three doping regimes in (a). The dashed lines in (c) and (e) mark $\beta_{\rm so}=\lambda_{\rm so}k_0$. Units are the same as in Fig.~\ref{fig2} of the main text.\label{fig1s}}
\end{figure*}

\begin{figure*}[h]
\includegraphics[width=0.75\textwidth]{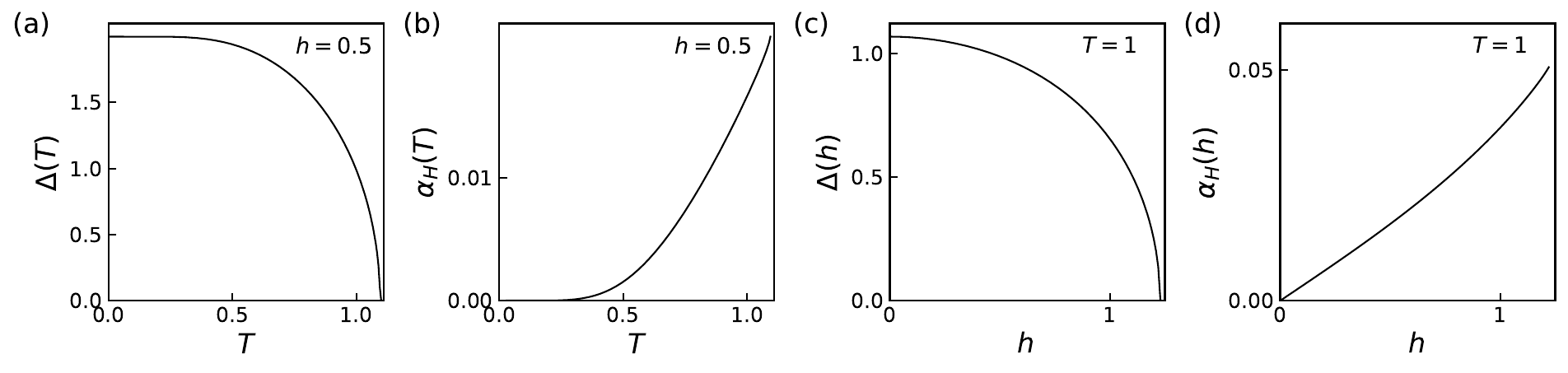}
\caption{(a) $\Delta(T)$ and (b) $\alpha_H(T)$ at $h=0.5$. (c) $\Delta(h)$ and (d) $\alpha_H(h)$ at $T=1$. Normal Hamiltonian parameters are $\lambda_{\rm so}k_0=6$, $\beta_{\rm so}=7$, and $t=500$, corresponding to point $a$ in Fig.~\ref{fig2}(b) of the main text. The superconducting gap $\Delta(h,T)$ is obtained by numerically solving Eq.~\ref{eq:Delta(h,T)}. \label{fig2s}}
\end{figure*}

In this section, we first discuss the conditions that maximize the Nernst signal for three different doping regimes of the s-wave pairing case, as indicated in Fig.~\ref{fig1s}(a,b). We then analyze the temperature and field dependence of the Nernst response $\alpha_H(h,T)$.

In the first regime ($\mu_1$, same as the main text but at a lower doping), we consider a fixed doping $\mu+|\beta_{\rm so}|=\text{const.}$ with Fermi energy lying in the spin-orbit gap but close to the lower band edge, where the contribution from the upper band is negligible. Fig.~\ref{fig1s}(c) shows that the hotspot of the Nernst signal shifts to the right (of the dashed line). Similarly, at higher dopings it shifts to the left. Those can also be found in Fig.~\ref{fig1s}(b). 

In the second regime ($\mu_2$), we consider the case where the Fermi energy lies above the spin–orbit gap with fixed doping $\mu-|\beta_{\rm so}|=\text{constant}$. Here, the Nernst signal is dominated by the upper band. As shown in Fig.~\ref{fig1s}(d), for low dopings ($\mu$ close to the upper band edge), enhanced thermal excitations near $k=0$ can produce relatively large $\alpha_H$. Since the energy dispersion around $k=0$ depends only weakly on the Ising and Rashba spin-orbit couplings, the resulting hotspot region of $\alpha_H$ is relatively insensitive to them.

In the third regime ($\mu_3$), we consider the case where the chemical potential lies at the midpoint of the two band edges, i.e., $\mu=0$, so that both the upper and lower bands contribute [see Fig.~\ref{fig1s}(e)]. Here, the Nernst signal originates from quasiparticles away from $k=0$, since at $k=0$ the energies are degenerate, leading to comparable particle and hole excitations that cancel each other. The Nernst signal remains maximized when the Ising and Rashba spin–orbit couplings are comparable [see the hotspot around $\beta_{\rm so} = \lambda_{\rm so} k_0 \approx 3$], but weakens at larger $\beta_{\rm so}$ due to suppressed quasiparticle excitations.

Finally, we discuss the behavior of $\alpha_H(T,h)$ for the $s$-wave pairing case. As shown in Fig.~\ref{fig2s}(a,b), the Nernst signal is exponentially suppressed at low temperatures, where the superconducting gap exceeds the thermal energy. With increasing temperature, thermally excited quasiparticles contribute to transport, leading to a monotonic enhancement of $\alpha_H$ up to $T=T_c$. Above the critical temperature, superconductivity is destroyed, and a finite quasiparticle Nernst response persists in the normal state.

The field dependence is shown in Fig.~\ref{fig2s}(c,d), where $\alpha_H(h)$ increases monotonically with $h$ within the superconducting regime, reflecting the enhanced time-reversal symmetry breaking induced by the Zeeman field. As a note, the calculation includes Rashba spin-orbit coupling but neglects orbital effects, so the critical field is not orbitally limited. Rashba coupling reduces paramagnetic depairing and allows the critical field to exceed the conventional Pauli limit, whereas orbital effects in real materials would suppress it.
\end{widetext}

\bibliography{ref}

\end{document}